\documentclass[prr,twocolumn,amsmath,amssymb,superscriptaddress,floatfix]{revtex4-2}%
\usepackage{graphicx}
\usepackage{dcolumn}
\usepackage{bm}
\usepackage{epsfig}
\usepackage{multirow}
\usepackage{rotating}
\usepackage{xcolor}

\begin{document}

\title{Spin liquid properties of the kagome material Cu$_3$(HOTP)$_2$}

\author{F. L.~Pratt}
\email{francis.pratt@stfc.ac.uk}
\affiliation{ISIS Neutron and Muon Source, STFC Rutherford Appleton Laboratory, Chilton, Didcot OX11 0QX, UK}
\author{D.~L\'{o}pez-Alcal\'{a}}
\affiliation{ICMol, Universitat Valencia, Catedr\'{a}tico Jos\'{e} Beltran Martinez, 46980 Paterna, Spain}
\author{V.~Garcia-Lopez}
\affiliation{ICMol, Universitat Valencia, Catedr\'{a}tico Jos\'{e} Beltran Martinez, 46980 Paterna, Spain}
\author{M.~Clemente-Le\'{o}n}
\affiliation{ICMol, Universitat Valencia, Catedr\'{a}tico Jos\'{e} Beltran Martinez, 46980 Paterna, Spain}
\author{J. J.~Baldov\'{i}}
\affiliation{ICMol, Universitat Valencia, Catedr\'{a}tico Jos\'{e} Beltran Martinez, 46980 Paterna, Spain}
\author{E.~Coronado}
\email{eugenio.coronado@uv.es}
\affiliation{ICMol, Universitat Valencia, Catedr\'{a}tico Jos\'{e} Beltran Martinez, 46980 Paterna, Spain}

\date{\today}

\begin{abstract}
The metal-organic-framework (MOF) compound Cu$_3$(HOTP)$_2$,
a.k.a. Cu$_3$(HHTP)$_2$,
 is a small-gap semiconductor containing a kagome lattice of antiferromagnetically coupled $S$=1/2  Cu$^\mathrm{II}$ spins with intra-layer nearest-neighbor exchange coupling 
$J \sim $ 2 K. 
The intra-layer $J$ value obtained from DFT+U calculations is shown to match with the experimental value for reasonable values of U.
Muon spin relaxation confirms no magnetic ordering down to 50~mK and sees spin fluctuations diffusing on a 2D lattice, consistent with a quantum spin liquid (QSL) ground state being present within highly decoupled kagome layers. 
Reduction of the spin diffusion rate on cooling from the paramagnetic region to the low-temperature QSL region reflects quantum entanglement. 
It is also found that the layers become more strongly decoupled in the low-temperature QSL region.
Comparison of results for the spin diffusion, magnetic susceptibility and specific heat in the QSL region suggests close proximity to a quantum critical point and a large density of low energy spinless electronic excitations. 
A Z$_2$-linear Dirac model for the spin excitations of the QSL is found to provide the best 
match with experiment. 
\end{abstract}

\maketitle

\newpage

\section{Introduction}

\label{introduction}

In the study of magnetic frustration, the kagome lattice has attracted particular attention, since it provides the most highly frustrated example of an antiferromagnet (AF) in a two-dimensional geometry.
Such frustration combined with the strong quantum fluctuations of a spin 1/2 magnetic unit is conducive to stabilizing a QSL ground state \cite{Zhou17,Savary17}.
This QSL state avoids the symmetry breaking associated with magnetic or valence bond order and has a dynamic and highly quantum-entangled character.
Entanglement is the primary QSL feature differentiating it from the high temperature paramagnetic (PM) phase.
For candidate QSL systems, some particular experimental challenges are detecting this entanglement and distinguishing the type of QSL from the many theoretical possibilities.
We demonstrate here the application of $\mu$SR for probing the degree of entanglement in a recently discovered kagome QSL system and apply the $\mu$SR results, in conjunction with previously reported specific heat and magnetic susceptibility, 
to determine characteristic parameters describing its QSL state. 

\begin{figure}[htb]
\includegraphics[width=\columnwidth]{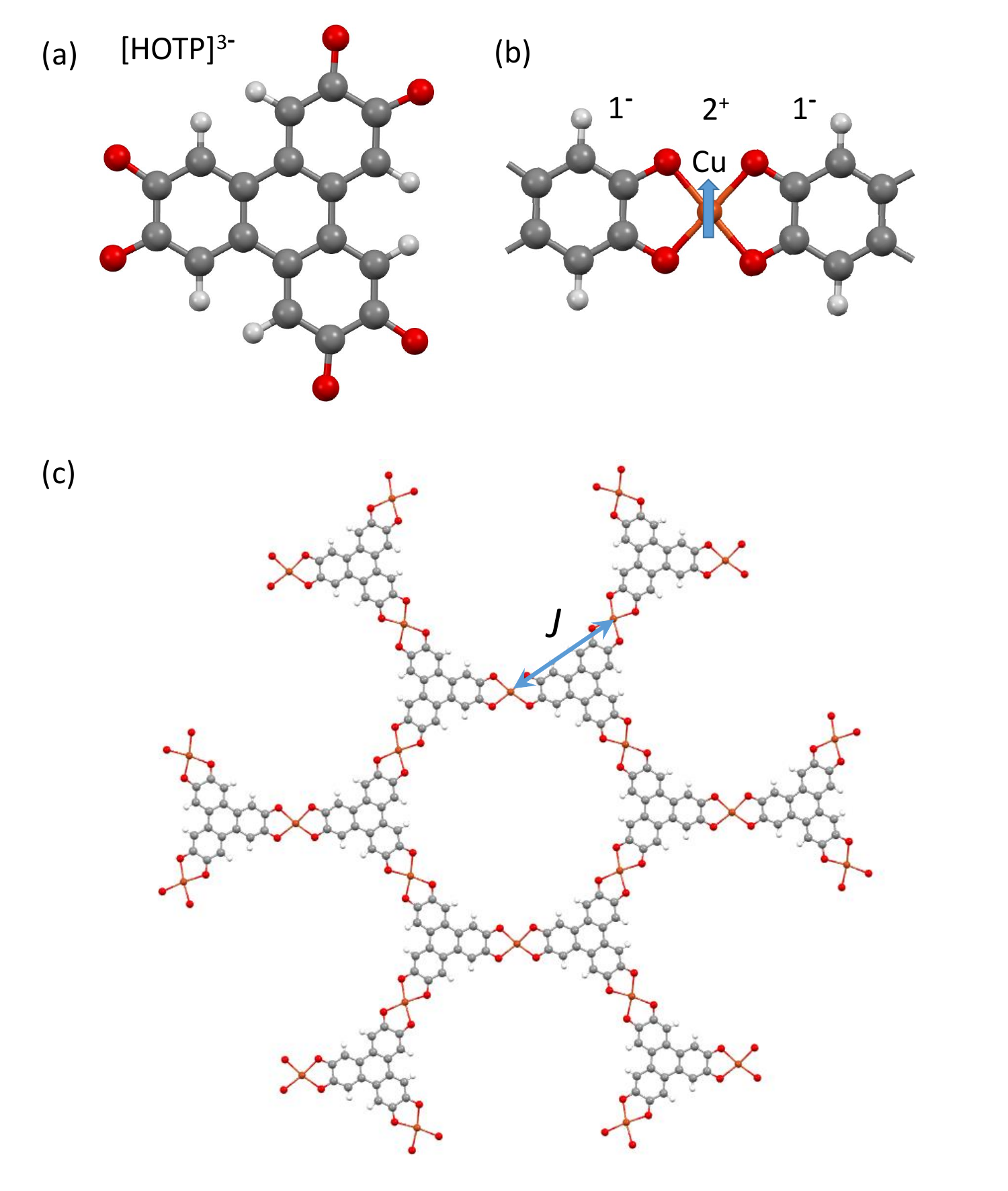}
\caption{
The kagome system Cu$_3$(HOTP)$_2$.
(a) The HOTP molecule. Oxygens are red, carbons are dark grey, hydrogens are light grey.
(b) An effective $S$=1/2 structural building block for the Cu$_3$(HOTP)$_2$ framework consisting of one Cu$^{2+}$ ion plus two oxyphenylene radical ligands, each ligand corresponding to one third of a full HOTP molecule.
A working model for the spin state of the structural unit has the radical ligands paired into a singlet state, leaving an unpaired spin 1/2 on the Cu ion.  
(c) The resulting $S$=1/2 kagome structure with nearest neighbour exchange coupling $J$.
}
\label{fig0}
\end{figure}

The theoretical QSL ground state of the $S$=1/2 Heisenberg AF model on the kagome lattice is still far from clear, despite numerous investigations. 
Some key questions are whether it is gapped or gapless and whether it has U(1) or Z$_2$ gauge structure.
Exact diagonalisation studies on finite lattices initially set a lower bound for the spin gap at 0.05~$J$ \cite{Waldtmann}, but more recent studies give a vanishingly small spin gap when extrapolated to the infinite lattice limit \cite{NakanoSakai11}.  
Density matrix renormalisation group studies on cylindrical lattice models initially found gapped states \cite{Jiang08,Yan11,Deppenbrock12,Nishimoto13}, but recent progress with the technique has revealed both gapless Dirac states \cite{He17,Zhu19} and a chiral gapped state \cite{Sun24}.
Variational Monte Carlo studies mostly find gapless U(1) Dirac states \cite{Ran07,Hermele08,Iqbal13,Iqbal14}, but the latest study \cite{ZhangLi20} finds a Z$_2$-linear Dirac state whose fluctuation spectrum is indistinguishable from that of the U(1) Dirac state. 
Tensor network studies have found either the U(1) Dirac state \cite{Liao17} or alternatively a Z$_2$ state with a vanishingly small gap \cite{Mei17}. 
A gapless Z$_2$ spinon Fermi surface (FS) state has also been found recently using a functional renormalization technique that goes beyond mean-field theory \cite{Hering19}.

On the experimental side, the most well-studied example of a kagome QSL to date is ZnCu$_3$(OH)$_6$Cl$_2$, known as herbertsmithite \cite{Shores05,Norman16}. 
Inelastic neutron studies show an excitation continuum that could reflect fractionalised spinons and set an upper limit to the gap of 0.015 $J$ \cite{Han12}.   
Experimental studies are complicated by a defect spin signal that can obscure the intrinsic properties of the kagome spins. 
This is due to site mixing between the Cu sites in the kagome plane and the interplanar Zn sites. 
These difficulties are illustrated by NMR studies from one group finding a gap of order 0.05 $J$ \cite{Fu15}, whereas a subsequent NMR study from another group was able to separate intrinsic and extrinsic contributions and find a gapless state \cite{Khuntia20}.

Finding kagome systems that avoid the defect spins arising from site mixing would provide a valuable contribution to the field and MOF systems can satisfy this aim.  
Kagome MOF systems made by combining transition metal M with the trigonal planar molecule hexahydroxytriphenylene (HHTP) were reported by Hmadeh and coworkers \cite{Hmadeh12} for M = Co, Ni, Cu. 
In the Co and Ni systems, extended M$_2$(HHTP)$_3$ kagome layers alternate with layers of isolated trinuclear M$_3$(HHTP) complexes.
For the case M~=~Cu, the whole system adopts the kagome structure \cite{Yang19} and the HHTP molecules are fully deprotonated to form hexaoxytriphenylene (HOTP). 
The kagome layer structure of the resulting MOF system Cu$_3$(HOTP)$_2$ \cite{naming} is shown in Fig.\ref{fig0}.
Although powder x-ray studies could not distinguish any regular stacking mode, a pattern of alternating displacements parallel to the layer planes was suggested as a structural model on the basis of previous DFT calculations \cite{Rubio18}. 
We show here that such a regular alternating structure is energetically unstable and any such periodically ordered stacking is frustrated by electrostatic interaction between next-nearest-neighbor layers.

The electronic, thermodynamic and magnetic characteristics of this material were reported recently \cite{Misumi20}.
Conductivity measurements show semiconducting behaviour with a charge energy gap of 0.34~eV.  
From fitting the magnetic susceptibility to a high temperature series expansion, the exchange coupling $J/k_\mathrm{B}$ was estimated to be 2.0~K.
A Curie-Weiss fit of the same data over a wide temperature range gave the Weiss constant $\theta=-3.4$~K. 
For coordination number $z$ = 4, this gives an estimate of the exchange coupling as $J/k_\mathrm{B} = 2\theta/z =$ 1.7 K, similar to that obtained from the high temperature expansion. 
No evidence for ordering was observed in measurements taken down to 38 mK and 
power law behaviour was observed for the non-phonon contribution to the specific heat, as well as the ac susceptibility \cite{Misumi20}.
The residual magnetic entropy at low temperatures was found to be 0.3 $R \ln$2, consistent with the highly degenerate ground state expected for Heisenberg spins on a kagome lattice \cite{ElstnerYoung94}.
All of these properties are consistent with a gapless QSL ground state.  

In this study we use first principles density functional theory to 
calculate the exchange coupling, the energetics of the layer stacking and the electronic structure for some simple stacking models, that can be compared against experiment.
For the experimental part of our study we use muon spin relaxation as a local probe of the magnetic properties down to 50 mK.
Spin dynamics is explored via the magnetic field dependence of the muon spin relaxation rate, which provides information about the spectral density of the spin fluctuations.  
These measurements are used to reveal a classical to quantum crossover in the spin dynamics and provide a gauge of the degree of quantum entanglement.
Taking the muon results alongside previously reported bulk magnetic and thermal properties allows some conclusions to be drawn about the nature of the low energy excitations in the QSL. 

\section{Methods}

\subsection{Synthesis}
\begin{figure}[h]
\includegraphics[width=0.95\columnwidth]{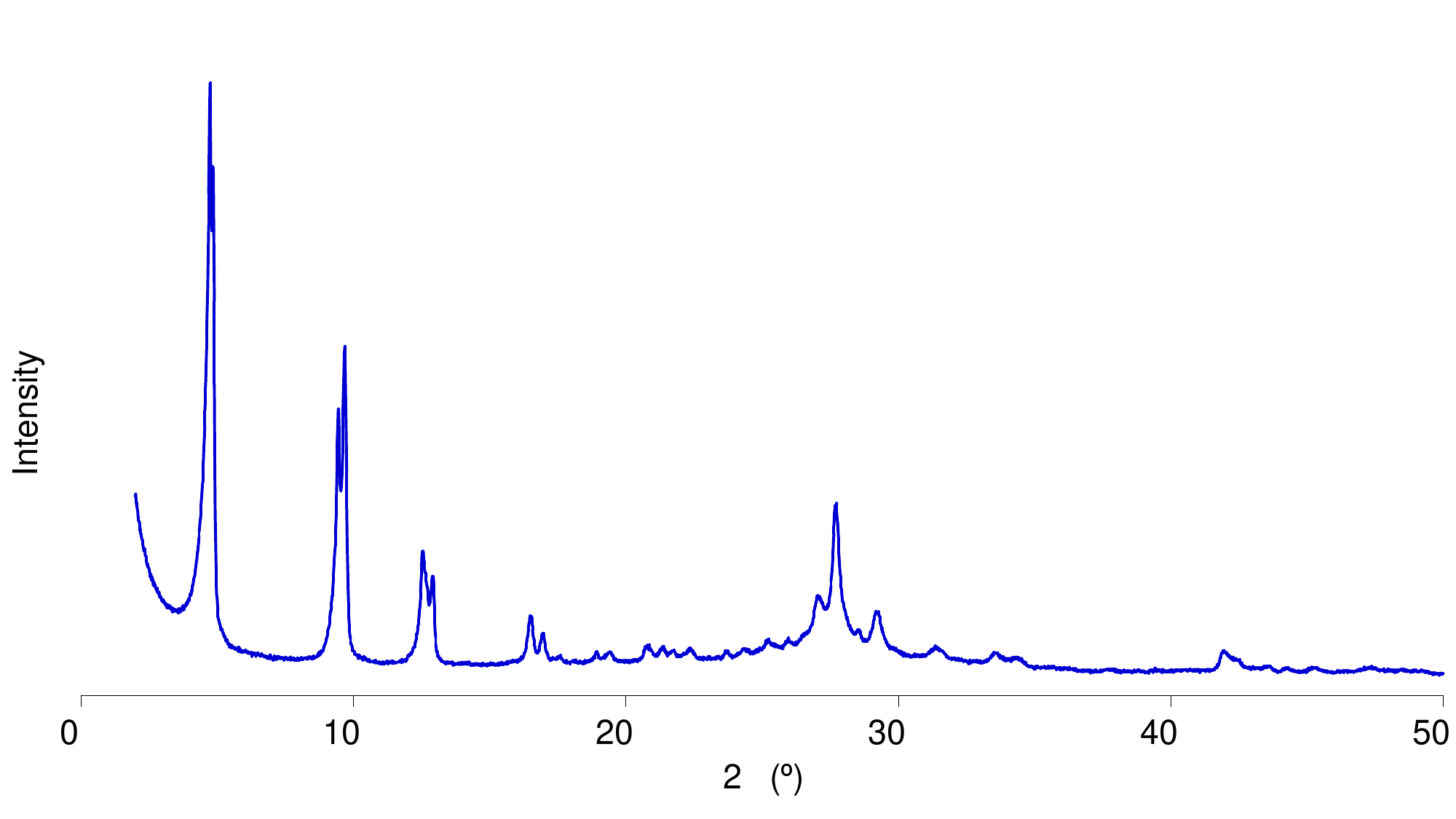}
\caption{
Powder x-ray pattern obtained for our sample of Cu$_3$(HOTP)$_2$.}
\label{fig-pxrd}
\end{figure}

Samples for the muon studies of Cu$_3$(HOTP)$_2$ that are reported here were prepared using the method of Hmadeh et al. \cite{Hmadeh12}.
200 mg of Cu(NO$_3$)$_2$·nH$_2$O, 140~mg of HHTP and 3 mL of N-methylpyrrolidone were mixed in 30 mL of deionized water in a hydrothermal Teflon autoclave. 
The mixture was heated at 85° C for 24 hours. The mixture was filtered and washed with H$_2$O and acetone. 
Elemental analysis found the composition C:~42.27~\%, H:~3.37~\%.
Calculated composition for Cu$_3$(C$_{18}$H$_6$O$_6$)$_2$(H$_2$O)$_{11}$ is C:~42.17~\%, H:~3.34~\%.
The powder x-ray pattern is shown in Fig.\ref{fig-pxrd} and is seen to be consistent with previous reports for this compound.

\subsection{Electronic properties}
We performed first principles spin-polarized density functional theory (DFT) calculations in the plane wave formalism as implemented in the Quantum ESPRESSO package \cite{Gianozzi09}. 
The exchange-correlation energy is calculated using the generalized gradient approximation (GGA) using the Perdew–Burke–Ernzerhof (PBE) functional \cite{Perdew96}. 
We selected standard solid-state pseudopotentials from the Materials Cloud database \cite{Prandini18}. 
The strong correlations of the $d$ electrons of the Cu ions of Cu$_3$HOTP$_2$ were considered by implementing the Hubbard-corrected DFT+U approach, 
choosing a standard on-site Hubbard U of 7.3 ~eV for the Cu atoms and 4~eV for C and O atoms in the simplified version proposed by Dudarev et al \cite{Dudarev98}. 
The electronic wave functions were expanded with well-converged kinetic energy cut-offs for the wave functions and charge density of 60 and 480 eV, respectively. 
The chemical structure of the bulk was fully optimized using the Broyden-Fletcher-Goldfarb-Shanno (BFGS) algorithm until the forces on each atom were smaller than $5{\times}10^{-3}$ Ry/au and the energy difference between two consecutive relaxation steps was less than $1{\times}10^{-4}$ Ry. 
We relaxed the molecular structure and the lattice parameters conserving the hexagonal lattice symmetry. 
Dispersion corrections were considered by applying semi-empirical Grimme-D2 corrections \cite{Grimme06}. 
The Brillouin zone was sampled by a fine $\Gamma$-centered 3 × 3 × 12 $k$-point Monkhorst–Pack mesh \cite{Monkhorst76} for SCF calculations and 6 × 6 × 16 for NSCF calculations.  

In the analysis of layer stacking energies, the slip translation and layer spacing from the slipped bilayer calculation were used as the starting point.
The highly planar monolayer structure was used for building the layer structure, starting with the slipped bilayer. 
The third layer was then added to this bilayer with the same layer spacing and slip distance, 
but varying the slip angle for the planar translation between second and third layers in 30$^\circ$ steps. 
A single point energy was evaluated for the three-layer structure versus slip angle. 
Since the O-Cu-O angles are not exactly 90$^\circ$, the actual slip angles differ a little from exact 30$^\circ$ steps. 
We corrected for this by averaging the values from positive and negative slip angles to give the expected symmetrical angular dependence versus nominal slip directions. 
The charge distribution was confirmed to be symmetrical about the Cu site, with charges of +1.27 on the Cu, -1.6 on the four O atoms and +1.3 on the four C atoms bonded to the O.

\subsection{Muon spectroscopy}
The muon measurements \cite{book} were carried out using the HiFi spectrometer at the ISIS Neutron and Muon Source. 
A helium dilution refrigerator provided sample temperatures down to 50~mK. 
The sample was 300~mg of powder mounted in a thin silver foil packet clamped onto a silver sample plate.
Some measurements were initially made in zero field to confirm the absence of magnetic ordering at low temperatures, 
but the bulk of the measurements were made in longitudinal field to provide information about the spin dynamics.
These LF measurements were made in fields supplied by the HiFi superconducting magnet, covering a range between 5~mT and 2.5~T.
Each individual measurement collected 3 $\times 10^7$ positron events resulting from the decay of the positive muons implanted into the sample. 
The data set is available from the ISIS Facility \cite{data}.

The hyperfine coupling expected for muon addition to the ligand was estimated by taking the molecule 1,2-dioxyphenylene as a simple model to represent one third of the HOTP molecule. 
For muonium addition to site 3 of this molecule we obtain a hyperfine coupling of 135~MHz from DFT using the Gaussian16 code \cite{Gaussian16} at the level of B3LYP/cc-pVDZ. 
This value has been corrected for quantum enhancement by calibration against the hyperfine coupling for the muoniated radical in benzene. The bare value before quantum correction was 110~MHz.

\section{Results}

\begin{figure}[htb]
\includegraphics[width=\columnwidth]{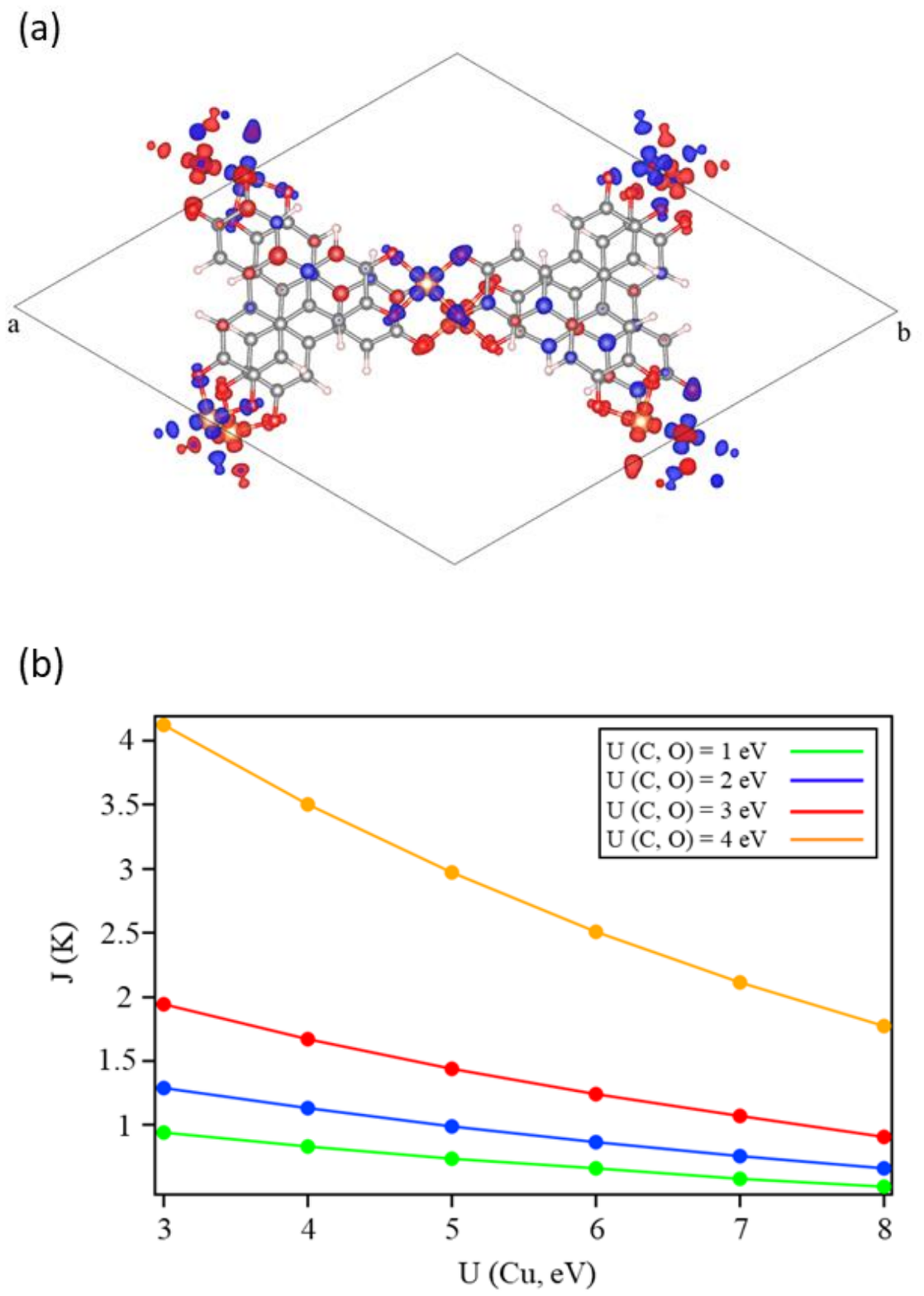}
\caption{
Calculation of exchange coupling.
(a) DFT+U calculated spin density of two consecutive layers of Cu$_3$(HOTP)$_2$. 
Color code: blue (red) spin up (spin down). Isosurface 0.009. 
(b) Magnetic exchange coupling calculated with different Hubbard U values. The experimental value of $J/k_\mathrm{B}$~=~2~K is matched for the combination U(Cu)~=~7.3~eV and {U(C,~O)~=~4~eV}.
}
\label{fig-spindenHubbard}
\end{figure}

\subsection{Electronic properties from DFT}

\label{DFTresults}

\subsubsection{Slipped versus eclipsed stacking}

\label{slipping-section}

As noted in section \ref{introduction}, the stacking of successive layers has been suggested to show a small displacement parallel to the layer planes. 
Based on this slipped arrangement, it was suggested that an ordered structure could have an alternating form \cite{Rubio18}. 
We first confirm this displacement via first principles calculations of a bilayer, showing that the fully eclipsed stacking (AA) is energetically unfavorable. 
The calculated stable arrangement of the bilayer exhibits a displacement of a 1.1 {\AA} between the layers towards the $b$ axis, imposing a slipped-parallel packing (AB) pattern on the structure.
This slipped-parallel stacking configuration is $\sim$1 eV more stable than the eclipsed one.  
In the relaxed structure we performed a bond-length analysis in order to elucidate the oxidation character of the HOTP$^{3-}$ ligand, 
because the C-O bond length could be indicative of the semiquinone oxidation state of the ligand unit \cite{Ovcharenko07}. 
We observed an average C-O bond length of 1.293 Å, which is indeed consistent with a semiquinone character for the HOTP$^{3-}$ ligand \cite{Meng22}.

\subsubsection{Magnetic ground state}

Our spin-polarized DFT+U calculations elucidate the magnetic properties of Cu$_3$(HOTP)$_2$, confirming the $S$~=~1/2 magnetic ground state. 
This is caused by the frustrated spin arrangement of the magnetic moments of the Cu$^{2+}$ atoms on the kagome lattice structure. 
Figure \ref{fig-spindenHubbard}(a) reveals that the spin density of the system is mainly located in $d_{x^2-y^2}$ orbital of the Cu$^\mathrm{II}$ atoms and in the $\sigma$ O-Cu bonds of each ligand molecule. 
The contributions to the spin density are different, since 79\% is present at each of the Cu$^\mathrm{II}$ atoms with the remainder on the ligand units (mainly located at the O atoms). 
Notice that the lack of spin density calculated for the HOTP$^{3-}$ radical (for which an unpaired electron delocalized over the $\pi$-system is expected) may be in agreement with the structural features of the solid. 
In fact, for similar MOFs based on the THQ ligand (THQ = tetrahydroxy-1,4-benzoquinone) this lack of magnetism was attributed to strong $\pi-\pi$ interactions between ligands of adjacent layers with eclipsed stacking, stabilizing spin singlet dimers \cite{Meng22}.

\subsubsection{Magnetic exchange coupling}

We followed a systematic broken symmetry approach to calculate the magnetic exchange coupling $J$, 
which can be estimated from the difference in energy of the possible spin configurations as $2J = E_{S=1/2} - E_{S=3/2}$. 
This broken symmetry method has been shown to generate accurate results for other trinuclear Cu systems \cite{Yoon05,Kawakami07,Spielberg10,Spielberg15,Baryshnikov17}. 
In these calculations we explored the effect of using different Hubbard U values until we reached convergence with the experimental value of $J/k_\mathrm{B}$ = 2 K, Fig.\ref{fig-spindenHubbard}(b). 

Magnetic coupling between adjacent layers is expected to be extremely small, as the spin density is entirely within the planar  $d_{x^2-y^2}$ and $\sigma$-electron orbitals and the $\pi$-electron and $d_{z^2}$ orbitals responsible for interlayer electronic interaction are orthogonal to these planar orbitals. 
In this case the magnetic coupling between the layers will be determined
by competition between AF exchange and ferromagnetic (FM) dipolar contributions of similar magnitudes.
The exchange contribution is calculated to be 0.13 K, whereas the dipolar contribution is calculated to be -0.07 K, giving an estimate of 0.06 K for the overall interlayer coupling.

\begin{figure}[b]
\includegraphics[width=\columnwidth]{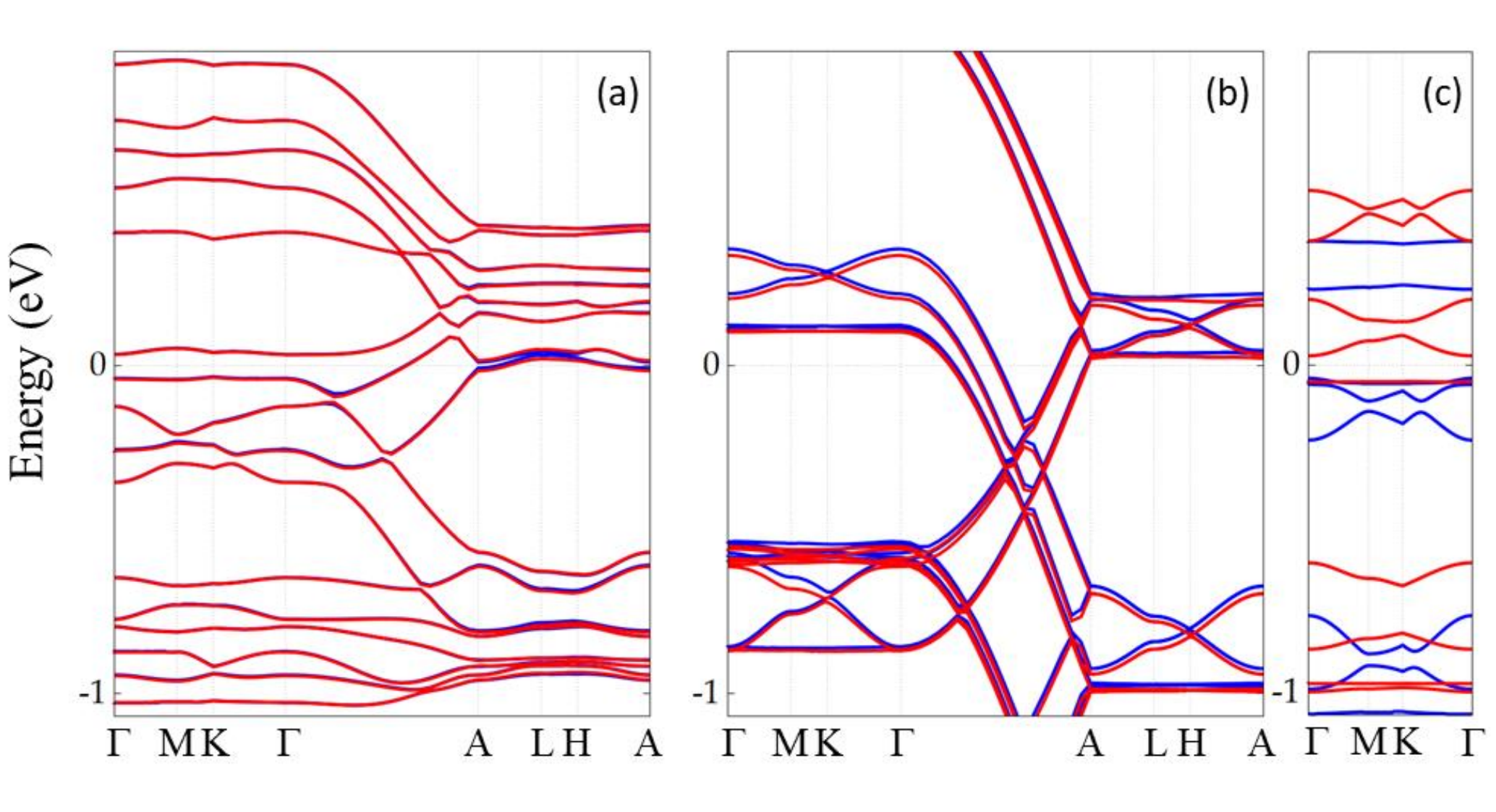}
\caption{
Band structures for previously proposed ordered stacking models.
(a) Calculated DFT+U band structure of Cu$_3$(HOTP)$_2$ for a regular zigzag slipped AB configuration. 
Color code: blue (red) spin up (spin down).
Corresponding band structures for (b) the unstable regular eclipsed AA configuration and (c) an isolated monolayer for comparison.
}
\label{fig-BS}
\end{figure}

\begin{figure*}[htb]
\includegraphics[width=1.6\columnwidth]{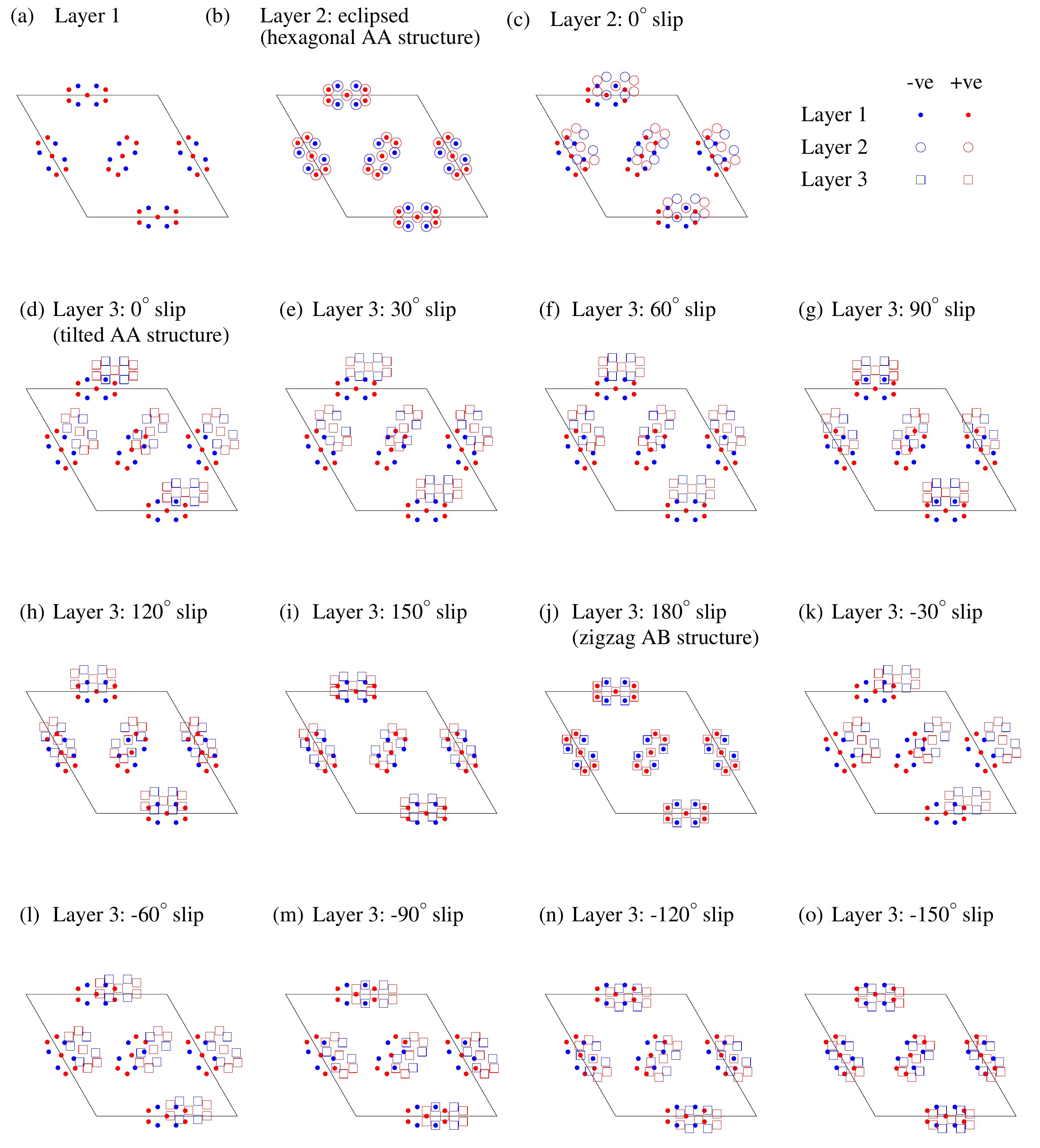}
\caption{
Illustration of the electrostatic frustration of the layer stacking. 
(a) The first layer. 
(b) Eclipsed stacking of a second layer is unstable. 
(c) Slipped stacking of a second layer is stable with 12 possible slip directions. 
(d) to (o) The 12 possible directions of slip for a third layer relative to the slip of the second layer. The charge alignment of layer 3 versus its next-nearest layer (layer 1) is shown.
Two possibilities for the lowest energy third layer slip direction are found at 120$^\circ$ and -120$^\circ$ relative to the slip direction of the second layer. 
}
\label{fig-stacking}
\end{figure*}

\subsubsection{Electronic properties of ordered stacking models}

Experimentally Cu$_3$(HOTP)$_2$ and similar kagome systems are shown to behave as semiconductors \cite{Rubio18}. 
However, theory for ordered structures usually predicts a metallic ground state for this type of lattice \cite{Chen15}. 
For Cu$_3$(HOTP)$_2$, the calculated band structure assuming a regular crystal formed on the basis of the AB slipped structure is shown Fig.\ref{fig-BS}(a). 
A continuous band gap of $\sim$0.25 eV is present on the plane of the layers ($\Gamma$-M-K) whereas a relatively wide dispersion is observed in the stacking direction ($\Gamma$-A-L-H), where a number of levels are crossing the Fermi energy. 
This indicates a strong interaction between metal-organic sheets due the $\pi-\pi$ interactions between the aromatic rings presents in the structure, which is in good agreement with the interlayer distance obtained of $\sim$3.15 Å. 
In order to elucidate the effect of this $\pi-\pi$ interaction on the band structure dispersion, we checked the atomic and orbital contributions on the projected density of states of the system. 
As can be seen in Fig \ref{fig-DOS1} in the Appendix, the contribution to the states near the Fermi energy is mainly due to the C and O atoms present in the system. 
Since our assumption is that the interactions in the stacking direction cause the observed dispersion in the band structure, 
one would expect that the orbitals involved in this interaction should be the ones pointing towards the stacking direction, i.e. the $p_z$ orbitals of C and O. 
This is exactly what is observed in Fig.\ref{fig-DOS2} in the Appendix, where the $p_z$ orbitals are the only orbitals that contribute to the DOS in this energy range. 
In order to further confirm the importance of $p_z$ overlap for the interlayer band width, we computed the band structure for Cu$_3$(HOTP)$_2$ but in an eclipsed AA configuration, 
where the layers are stacked perfectly one on top of one another, Fig.\ref{fig-BS}(b). 
Here we can see that the dispersion in the $\Gamma$-A-L-H direction has increased substantially, since this arrangement causes the maximum possible overlap between the $p_z$ orbitals of the C and O atoms in the organic ligands.

\subsubsection{Origin of the semiconducting properties}

Since the experimental results indicate a highly 2D semiconducting state, rather than a quasi-1D metallic state, there is clearly a major inconsistency between 
these theoretical calculations for simple regular stacking scenarios and experiment. 
This issue has been widely discussed, reaching the conclusion that the role of defects in the arrangement of stacking units is likely to limit coherent interlayer transport and prevent large metallic conductance in this class of materials \cite{Foster16,Foster18,Mancuso20}.
In the extreme limit, the system could be represented by non-interacting monolayers, with the corresponding band structure shown in Fig.\ref{fig-BS}(c). 
This shows a small semiconducting energy gap that is comparable to that observed experimentally. 
The discrepancy here between crystalline models and experiment can be further resolved by noting that there is actually no experimental evidence for any regular mode of interlayer stacking.
In the next section we model the energetics of interlayer stacking driven by the intralayer charge distribution to provide some insight into this aspect of the system.

\subsection{Model for disordered layer stacking}

\begin{figure}[tb]
\includegraphics[width=0.85\columnwidth]{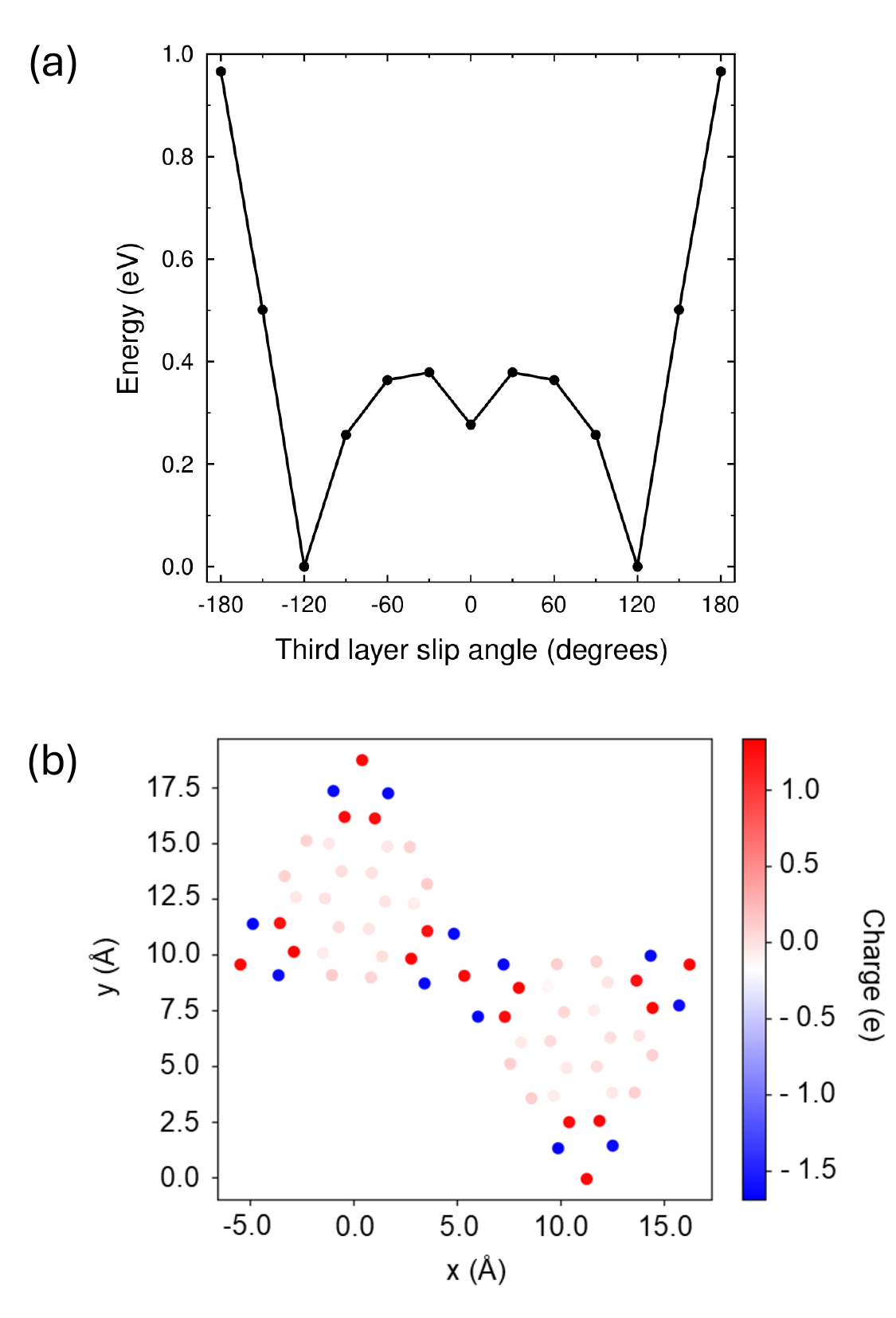}
\vspace{-5mm}
\caption{
Stability of different next-nearest-layer configurations.
(a) Relative energy versus the slip angle of the third layer. 
(b) Model charge distribution within the unit cell of a layer.
}
\label{slip-fig}
\end{figure}

A simple electrostatic model that can produce disorder in the layer stacking pattern of this type of system is illustrated in Fig.\ref{fig-stacking}. 
Taking the first layer, Fig.\ref{fig-stacking}(a), as a reference point, eclipsed stacking of a second layer, Fig.\ref{fig-stacking}(b), is not stable in comparison with slipped stacking, Fig.\ref{fig-stacking}(c), that primarily aligns positive charge on one Cu with negative charge on one of the ligand oxygens. 
This picture based on electrostatic alignment is consistent with the DFT results of section \ref{DFTresults}.
There are 12 possible directions for this slip with equal stabilisation energy, we pick one of these for the second layer and define this as our reference direction 0$^\circ$, Fig.\ref{fig-stacking}(c).  
For the third layer there are again twelve possible slip directions, which each have the same stabilisation energy with respect to nearest layer interaction.
Selection of the layer stacking configuration then depends on differences in energy at the level of the next-nearest layer interaction, which is illustrated for the twelve possibilities in Fig.\ref{fig-stacking}(d) to Fig.\ref{fig-stacking}(o).
From the charge alignment patterns, it can be seen that the directions +120$^\circ$, Fig.\ref{fig-stacking}(h), and -120$^\circ$, Fig.\ref{fig-stacking}(n), are the most stable, giving two energetically equivalent choices for adding the new layer. 

This picture is backed up by model calculation of the relative energies versus slip angle for the third layer, as shown in Fig.\ref{slip-fig}(a) for the charge distribution model shown in Fig.\ref{slip-fig}(b). 
It can be seen from Fig.\ref{slip-fig}(a) that the zigzag AB mode with 180$^\circ$ slip angle is highly unstable compared to the +120$^\circ$ and -120$^\circ$ slip angles. 
With these two stable slip angles, a new binary choice of slip direction is then found for each new layer, leading to an intrinsic frustration-driven disordering of the stacking pattern for this type of structure. 

This modelling shows a likely reason for the lack of any evidence for a regular stacking pattern in the powder x-ray data.
The effect of stacking disorder in the closely related compound Ni$_3$(HITP)$_2$ was very clearly shown to switch the band structure from metallic to semiconducting \cite{Foster18}.  
The mechanism we have identified here for disordered stacking in Cu$_3$(HOTP)$_2$ therefore points towards the origin of the effective two-dimensional semiconducting character of its measured electronic properties. 

\subsection{Muon spin relaxation}

We have previously used zero-field (ZF) and longitudinal-field (LF) $\mu$SR to study spin dynamics in a range of low-dimensional spin 1/2 quantum systems, 
e.g. Heisenberg AF spin chains \cite{deocc,1DHAF2}, spin ladders \cite{ladders} and layered quantum spin liquids \cite{TaS2,YbZnGaO4}.  
In all of these studies the first step is to use ZF measurements to confirm that magnetic ordering is suppressed and the second step is to use LF studies to probe the the shape of the spectral density function describing the distribution of the fluctuations versus frequency, with the probe frequency being proportional to the applied field.

\subsubsection{ZF and LF $\mu$SR data}

\begin{figure}[tb]
\includegraphics[width=\columnwidth]{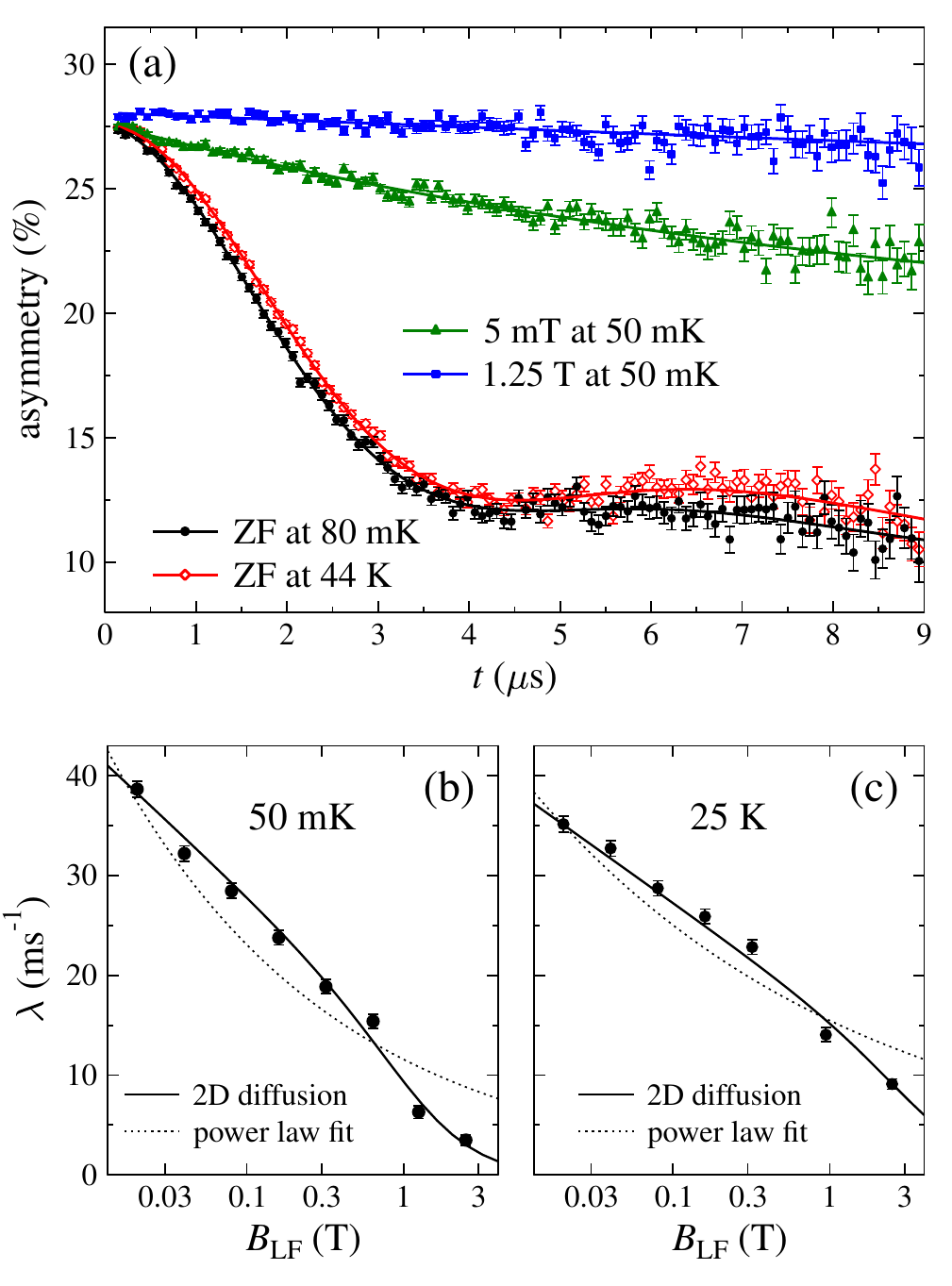}
\caption{
Muon spin relaxation.
(a) Example data 
showing ZF muon spin relaxation at 80~mK and 44~K and LF muon spin relaxation at 50 mK for fields of 5~mT and 1.25~T, where the relaxation is purely electronic in origin.
(b,c) Muon spin relaxation rate versus $B_\mathrm{LF}$ compared for two different temperatures, one well below the exchange energy $J$ and the other well above. 
The solid lines show fits to the 2D spin diffusion model of Eq.(\ref{relaxation}). 
The dotted lines show the best attempts at fitting to power laws, which provide a much poorer description of the data than the 2D diffusion fits. 
}
\label{fig2}
\end{figure}

Examples of ZF and LF $\mu$SR data for Cu$_3$(HOTP)$_2$ are shown in Fig.\ref{fig2}(a). 
No evidence for magnetic ordering is found in our studies taken down to 50 mK, consistent with the 38 mK upper limit for ordering reported earlier \cite{Misumi20}.
The ZF relaxation shows primarily a Kubo-Toyabe form at all temperatures, reflecting a dominant contribution from static nuclear dipolar fields,
along with a weak electronic contribution resulting from spin dynamics, which is present at all temperatures.
Application of moderate LF quenches the nuclear contribution to the relaxation, leaving just the electronic contribution, which has the slow Lorentzian form that is expected for fast electron spin dynamics.
The electronic relaxation rate decreases with increasing field, reflecting the characteristic frequency-dependent spectral density of the spin dynamics.
It can be seen that a significant relaxation can still be measured even for fields above 1~T, Fig.\ref{fig2}(a), indicating the presence of a rapid form of spin dynamics that is not easily quenched by the field.

\subsubsection{2D spin diffusion}

For spins diffusing in a 2D layer, the spin autocorrelation function $S_\mathrm{2D}(t)$ reflects a random walk process describing the propagating spin excitations that takes the form \cite{Butler76}
\begin{equation} 
S_{2D}(t) = [\exp(-2D_\mathrm{2D}t) I_0 (2D_\mathrm{2D}t)]^2,
\end{equation}
where $I_0$ is a Bessel function and $D_\mathrm{2D}$ is the diffusion rate in the layer.
The corresponding spectral density $J_\mathrm{2D}(\omega)$ is then given by the Fourier transform of $S_\mathrm{2D}(t)$.  
For an implanted muon probe linked to a spin site in the lattice via a contact hyperfine coupling $A$ the muon spin relaxation rate is given by
\begin{equation}
\lambda(B) = \frac{A^2}{4} J_\mathrm{2D}(\gamma_e B_\mathrm{LF}),
\label{relaxation}
\end{equation}
where $\gamma_e/ 2\pi$ = 28.02 GHz T$^{-1}$ is the gyromagnetic ratio of the electron and $B_\mathrm{LF}$ is the magnetic field.

\begin{figure}[tb]
\includegraphics[width=\columnwidth]{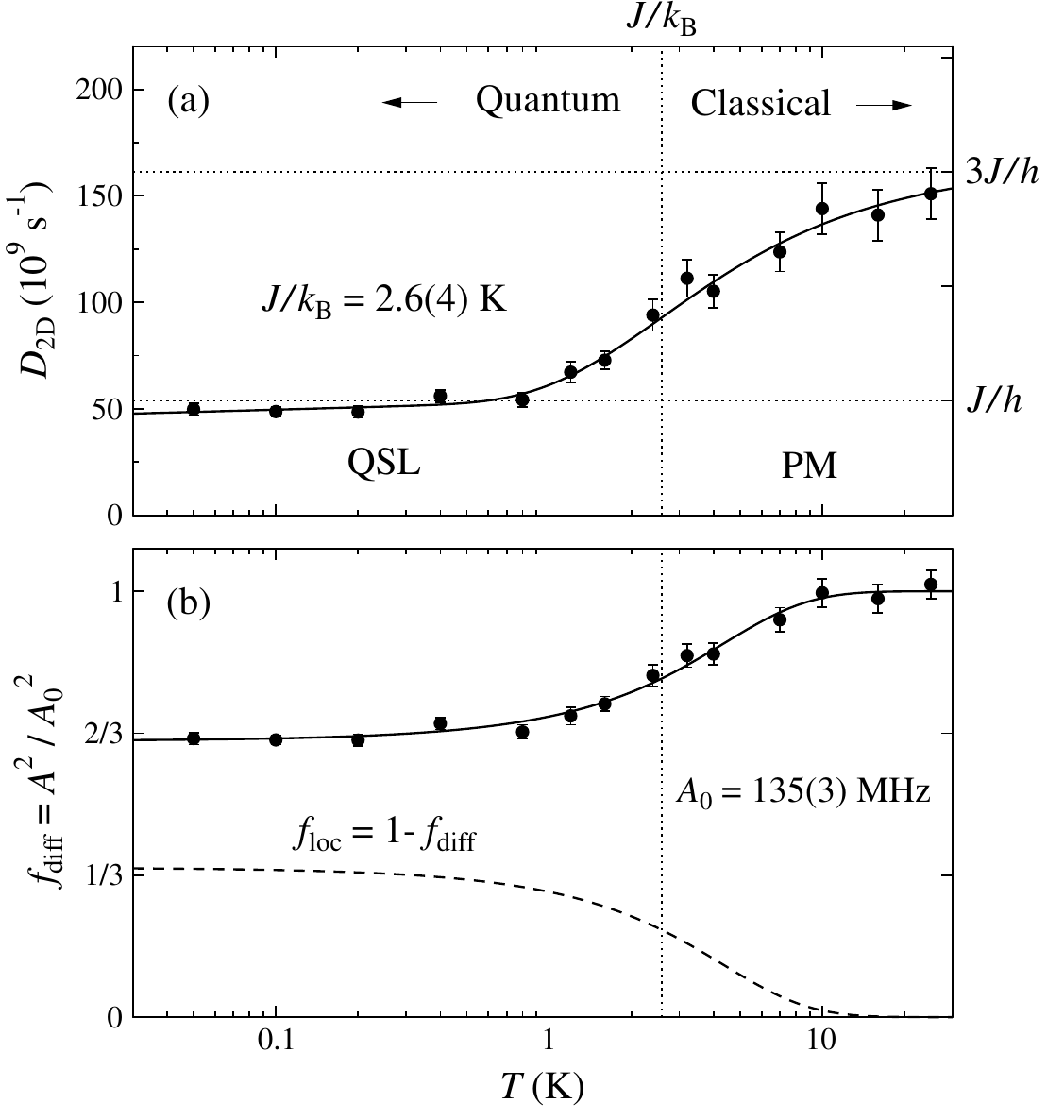}
\caption{
Classical to quantum crossover.
(a) Temperature dependence of the 2D spin diffusion rate, $D_\mathrm{2D}$, spanning between the QSL and PM regimes. 
The solid line is a fit to Eq.(\ref{Dfit}). 
(b) Temperature dependence of the diffusive fraction. 
The hyperfine coupling $A_\mathrm{0}$ has been derived from the high temperature PM regime where the excitations are fully diffusive.
In the QSL regime the diffusive fraction is reduced to 2/3, suggesting localization of 1/3 of the excitations.
}
\label{fig3}
\end{figure}

Example plots of the relaxation rate versus field are shown in Fig.\ref{fig2}(b,c), along with fits to Eq.(\ref{relaxation}).
The two parameters obtained from the fits are the 2D diffusion rate, $D_\mathrm{2D}$, derived from the field scaling of the $J_\mathrm{2D}$ function and the effective value of $A^2$ derived from the amplitude scaling factor for $\lambda$, as described by Eq.(\ref{relaxation}).  It can be seen from Fig.\ref{fig2}(b,c) that the 2D diffusion model provides a good representation of the observed field dependence.  
In contrast, a power law field dependence associated with a 1D diffusion model \cite{deocc,1DHAF2} is not found to be compatible with the data.

\subsubsection{Interlayer coupling}

The 2D model can be extended to allow for the effects of slow spin diffusion in the interlayer direction $D_\perp$.  At 25~K the ratio $D_\mathrm{2D}/D_\perp$ is then estimated to be 1.0(4)$\times$10$^2$ and at 50~mK the lower limit for the ratio is estimated to be 10$^{4}$. 
This gives an experimental estimate of the effective interlayer coupling at higher temperature of 0.02(1) K.
As noted in section \ref{DFTresults}, the dipolar contribution is -0.07 K, so the experimental estimate for the exchange contribution to the interlayer coupling is 0.09(1) K, which is reasonably consistent with the value of 0.13 K obtained from the broken symmetry calculation in section \ref{DFTresults}. 

The increased anisotropy ratio at low temperature 
points towards the presence of a highly effective dimensional reduction mechanism in the quantum region.
The origin of this strong dimensional reduction will be addressed in section \ref{Discussion}.

\subsubsection{Quantum to classical crossover}
The $T$ dependence of the fitted parameters obtained from the 2D diffusion model is shown in Fig.\ref{fig3}.
From Fig.\ref{fig3}(a) it can be seen that the diffusion rate is almost constant in the QSL region where $T \ll J/k_\mathrm{B}$ and increases by about a factor of three in the PM region where $T \gg J/k_\mathrm{b}$.  
The $T$ dependence can be fitted to the sum of a term with a weak power law $n$ and an activated term reflecting the thermal excitation of singlets
\begin{equation}
D_\mathrm{2D}(T) = D_\mathrm{0} {T^n} + D_\mathrm{1} \exp(-J/T).
\label{Dfit} 
\end{equation}
The values obtained from this fit are $D_\mathrm{0}$ =  53(4) $\times$ 10$^{9}$~s$^{-1}$, $n$ = 0.03(3), $D_\mathrm{1}$~=~103(11)$~\times$~10$^{9}$~s$^{-1}$ and $J/k_\mathrm{B}$ =  2.6(4) K. 
The fittted $J$ value is broadly comparable to that obtained from the susceptibility \cite{Misumi20}.  
It is found that the value of $D_\mathrm{0}$ closely matches 
$J/h$ = 54 $\times$ 10$^{9}$ s$^{-1}$ and $D_\mathrm{1}$ closely matches 2$D_\mathrm{0}$.
High $T$ saturation of $D_\mathrm{2D}$ starts around 10 K and at this point the magnetic entropy is expected to be close to $R \ln 2$. 
The magnetic specific heat shows similar behaviour, with the entropy change becoming saturated in the region above 7 K  \cite{Misumi20}.
The crossover from the low-$T$ quantum region to the high-$T$ classical PM region is qualitatively similar to that recently observed in the triangular lattice QSL system YbZnGaO$_4$ \cite{YbZnGaO4}. 
However there are some notable differences in the scaling between the asymptotic values of $D_\mathrm{2D}$ and $J/h$.  
Here the QSL region is the one that closely matches $J/h$, whereas for YbZnGaO$_4$ it was the PM region that matched $J/h$. 
The ratio of $D_\mathrm{2D}$ values between PM and QSL regions is 3 in this case, whereas it was more than 10 for YbZnGaO$_4$ \cite{YbZnGaO4}.
This contrasting behaviour may reflect the different underlying lattices and differences in the quantum entanglement lengths in the respective QSL phases.
The quantum entanglement will be discussed in a later section. 

\subsubsection{Localized states}
The normalised $T$ dependence of the $A^2$ coupling factor in Eq.(\ref{relaxation}) is shown in Fig.\ref{fig3}(b). 
In the paramagnetic region ($T \gg J/k_\mathrm{B}$), the value of $A$ is found to be $A_0$ = 135(3) MHz, consistent with the size of hyperfine coupling expected in a muoniated molecular radical probe state formed on the ligand (see Methods).  
In the QSL region ($T \ll J/k_\mathrm{B}$), the effective value of $A^2$ falls to around 2/3 of its high $T$ value. 
We do not expect strong $T$ dependence in $A$, so this drop in effective scaling factor is assigned to a reduced fraction of diffusive excitations in the overall spectral density, i.e. $f_\mathrm{diff} = A^2/A_0^2$. 
The low temperature missing fraction in the spectral density is assigned to a significant fraction of low frequency localized excitations $f_\mathrm{loc} = 1 - f_\mathrm{diff}$ that is suggested to emerge in the QSL region of Fig.\ref{fig3}(b). 
The spectral region for finding these localised excitations is below 0.02~T. 
At 50~mK the low field relaxation can be modelled as the sum of the 2D diffusion term, obtained from fitting to the higher field data, and a Lorenzian term with a low-field cut-off. 
From the cut-off field, the characteristic fluctuation rate of the localized excitation is estimated to be around 80~MHz at 50~mK. 
As expected from the $f_{\mathrm{loc}}$ inferred from Fig.\ref{fig3}(b), the low-field Lorenzian term is absent from the data at 10~K and above.

\begin{figure}[htb]
\includegraphics[width=\columnwidth]{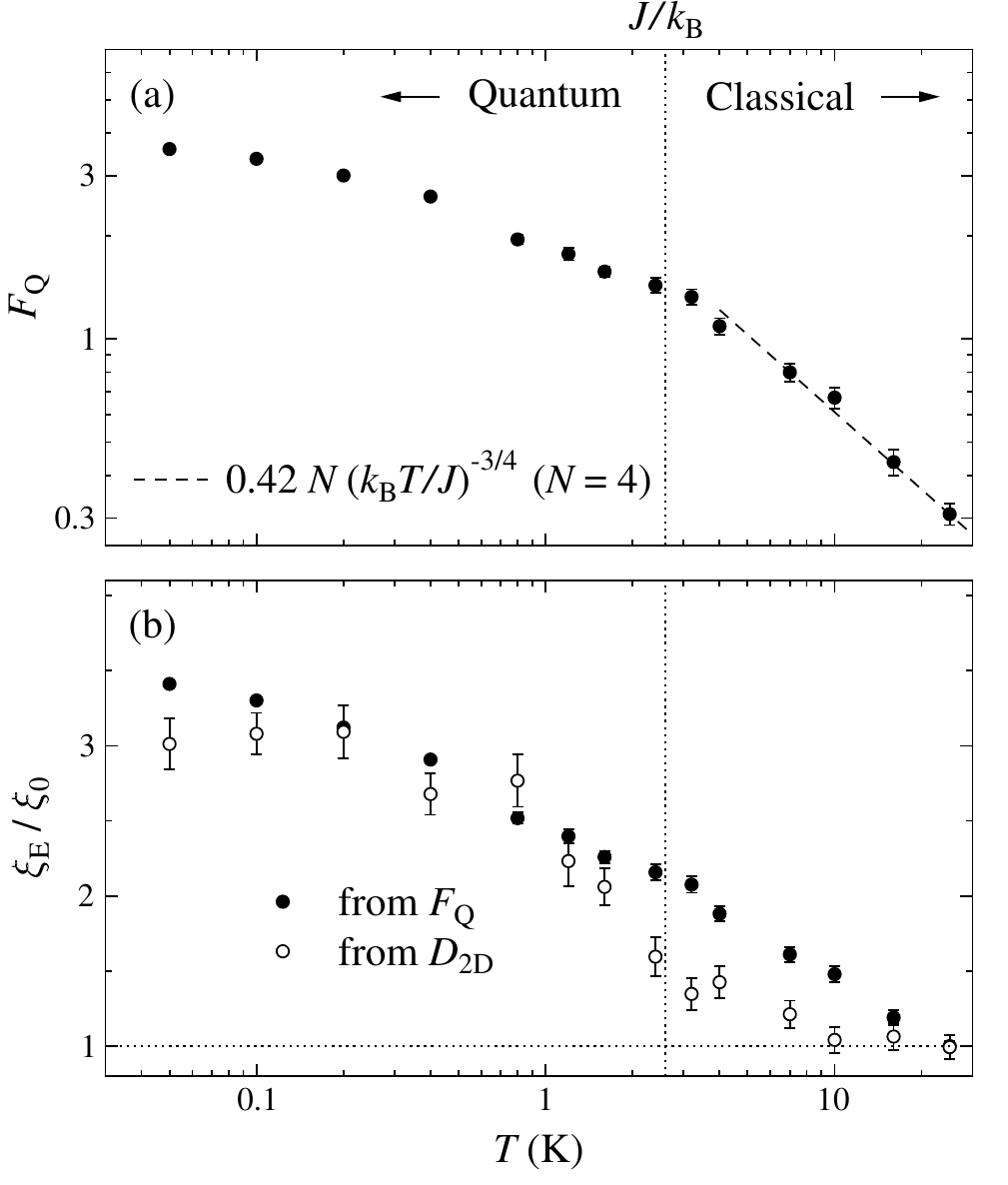}
\caption{
Quantum entanglement.
(a) The parameter $F_\mathrm{Q}$ derived from $J_\mathrm{2D}$ using Eq.\ref{FQ}. 
The $T$-dependence in the high $T$ region is consistent with the universal $T^{-0.75}$ power law for $N$~=~4 entangled particles \cite{Hauke16} (dashed line). 
(b) Entanglement length $\xi_E$ normalised to $\xi_0$ (the value at 25 K).  Estimates of this property as derived from $F_\mathrm{Q}$ and $D_\mathrm{2D}$ are compared. 
} 
\label{fig4}
\vspace{0mm}
\end{figure}  

\subsubsection{Quantum entanglement}

\label{Entanglement}
The 2D diffusion rate $D_\mathrm{2D}$ is related to the diffusion constant, $D$, the spinon velocity, $v$, momentum relaxation time, $\tau$, and the hopping distance or mean free path, $l=v\tau$, via
\begin{equation}
D = \frac{1}{2} v^2 \tau = \frac{1}{2} v l = D_\mathrm{2D} l^2,
\end{equation}
so that
\begin{equation}
D_\mathrm{2D} = \frac{v}{2 l} = \frac{1}{2\tau}.
\label{D2D}
\end{equation}

If the velocity is independent of $T$, as would be expected for a linear Dirac spinon dispersion or a spinon Fermi surface (FS) state, then $D_\mathrm{2D}$ follows the inverse of the mean free path $l$. 
Whereas $l$ is expected to match the nearest neighbour inter-site distance $a$ in the classical region, 
in the QSL region the quantum entanglement is expected to increase the length scale for the diffusion steps.
In this case the $l$ value derived from Eq.(\ref{D2D}) can be taken to give a measure of the entanglement length $\xi_E$ \cite{Verstraete04}. 
Another way of quantifying the entanglement is to use the quantum Fisher information metric 
$F_\mathrm{Q}$ \cite{Hauke16,Laurel21}, defined here in terms of the spectral density
\begin{equation}
F_\mathrm{Q} = \frac{4}{\pi} \int_0^{\infty} \tanh^2\left(\frac{\hbar\omega}{2k_\mathrm{B}T}\right) J(\omega) d\omega, 
\label{FQ}
\end{equation}
which is shown in Fig.\ref{fig4}(a).
The high $T$ behaviour is consistent with the predicted universal $T^{-3/4}$ form \cite{Hauke16}, whereas a weaker $T$-dependence is found for $T < J/k_\mathrm{B}$.
Since the number of locally entangled spins within a 2D layer can be taken to scale with $F_\mathrm{Q}$, another estimate of the entanglement length is given by $F_\mathrm{Q}^{1/2}$. 
The $T$-dependences of the normalised $\xi_E$ values calculated via $F_\mathrm{Q}$ and $D_\mathrm{2D}$ are compared in Fig.\ref{fig4}(b).
Both increase approximately threefold in the quantum region compared to the classical region. 
It is notable that the $D_\mathrm{2D}$-derived value shows a more well-defined transition around $J$, 
whereas the value from $F_\mathrm{Q}$ changes more gradually.

\section{Discussion}

\label{Discussion}

\subsection{Low-energy excitations of the QSL}

\renewcommand{\arraystretch}{1.1}
\renewcommand{\tabcolsep}{9.1mm}
\begin{table*} [htb]
\centering
\caption{QSL excitation models, with predicted parameters compared against experimental values of $n_\mathrm{D}$ from $\mu$SR,  
 $n_\mathrm{\chi}$ from susceptibility and $n_\mathrm{C}$ from specific heat via Eqs.(\ref{nD}), (\ref{nChi}) and (\ref{nC}). 
A spinon FS (line 1) is not a good overall fit with experiment (line 7), even with the inclusion of gauge fluctuations \cite{YbZnGaO4} (GF, line 2). 
Spinons with quadratic dispersion (line 3) can be excluded, since their predicted $n_\mathrm{C}$ and $n_\mathrm{\chi}$ values are both inconsistent with experiment.
Spinons with the quartic dispersion q~=~-0.5, could match both exponents $n_\mathrm{D}$ and $n_\mathrm{C}$ (line 4), 
but the predicted $n_\mathrm{\chi}$ would not be consistent with experiment.
In contrast, linearly dispersing spinons are consistent with experiment for both $n_\mathrm{D}$ and $n_\mathrm{\chi}$ (line 5). 
In this case the observed low value of $n_\mathrm{C}$ can be assigned to dominant singlet to singlet
excitations with quadratic dispersion and $\nu$ close to the mean field value of 0.5  (line 6).
Best assignments between excitation model and experiment are indicated in bold.
 \label{table1}
 }
\vspace{2mm}
\begin{tabular}{lccccc}
\hline\hline
 & $q$ & $\nu$ & $n_\mathrm{D}$  & $n_\mathrm{\chi}$ & $n_\mathrm{C}$ \\
\hline
1. Spinon (FS) & 0 & & 0 & 0 & 1 \\
2. Spinon (FS+GF) & 0 & & 0.33 &  & 0.67 \\
3. Spinon (quadratic) & 0 & 0.673 & 0.03 & -1.31 & 1.02   \\
4. Spinon (quartic) & -0.5 & 0.673 &  0.03 & -1.81 &  0.52 \\
{\bf 5. Spinon (linear)} & 1 & 0.673 & {\bf 0.03} & {\bf -0.31} & 2.02  \\
{\bf 6. Singlet (quadratic)} & 0 & 0.51 & & & {\bf 0.52} \\
{\bf 7. Experiment} & & & {\bf 0.03} & {\bf -0.31} & {\bf 0.52} \\
\hline\hline
\end{tabular}
\end{table*}

\subsubsection{Power laws and critical exponents}

The weak $T$-dependence of $D_\mathrm{2D}$ in the QSL region contrasts with the power law dependence reflecting quantum critical (QC) fluctuations that was recently found in $D_\mathrm{2D}$ for the triangular lattice QSL system 1T-TaS$_2$ \cite{TaS2}.
This might be taken to indicate only weak QC fluctuations here. 
However the power law for the electronic specific heat in  Cu$_3$(HOTP)$_2$ was recently reported to be $n_\mathrm{C}$~=~0.52 \cite{Misumi20}, which can be compared to the 2/3 power law previously found for herbertsmithite \cite{Helton07}.
This value suggests the presence of significant fluctuations, as it is substantially reduced from the expected value in the absence of such fluctuations, which would either be 1 for a spinon FS state or 2 for a Dirac QSL. 
A low temperature power law for the magnetic susceptibility was also reported as $n_\mathrm{\chi}$~=~-0.31 for $T$ in the range 0.04 to 0.3~K \cite{Misumi20}. 
This differs from the expected value of 0 for a spinon FS without fluctuations.
In the QC scenario, the power law versus $T$ for the momentum relaxation time is $2/\nu-3$ \cite{Qi09}, where $\nu$ is the correlation length critical exponent. 
From Eq.(\ref{D2D}) we expect the diffusion rate power law to be
\begin{equation}
n_\mathrm{D} = 3-2/\nu.
\label{nD}
\end{equation} 
For the first term in Eq.(\ref{Dfit}) that determines the $D_\mathrm{2D}$ dependence below 1~K in  Fig.\ref{fig3}(a), we obtained the relatively weak exponent $n_\mathrm{D}$~=~0.03(3), leading to the estimate $\nu$~=~0.673(7).
This is consistent with the value of 0.672 expected for the O(2) criticality class \cite{DePolsi20} and notably lower than the O(4) value of 0.748 \cite{DePolsi20} that was found to be consistent with both $n_\mathrm{D}$ and $n_\mathrm{C}$ in the case of the triangular lattice QSL 1T-TaS$_2$ \cite{TaS2}.

\begin{figure}[hbt]
\includegraphics[width=\columnwidth]{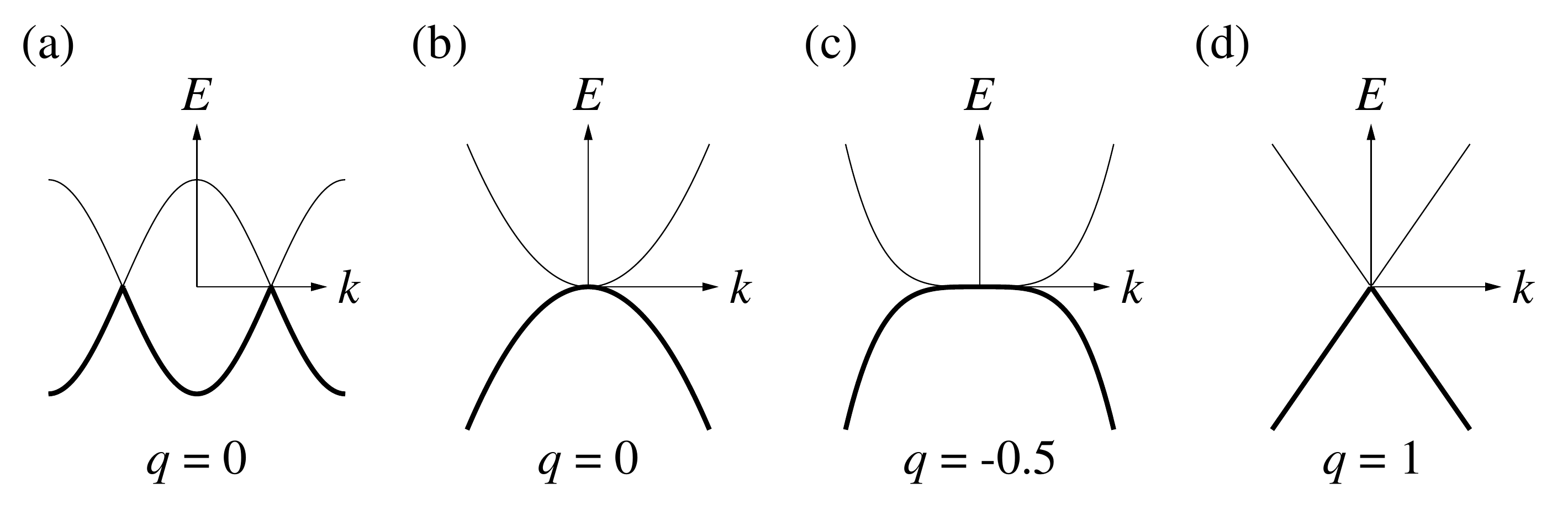}
\caption{
Types of spinon dispersion for 2D gapless QSL states. (a) Large area spinon FS and (b-d) band-touching states having (b) quadratic, (c) quartic and (d) linear dispersion.
}
\label{QSLstates}
\end{figure}

The predicted value for $n_\mathrm{\chi}$ is
\begin{equation}
n_\mathrm{\chi}  =  q - \gamma =  q - (2-\eta)\nu
\label{nChi}
\end{equation}
where $q$ is the power law for the density of states (DOS) versus energy and $\eta \sim 0.04$ \cite{DePolsi20}.
The value of $n_\mathrm{C}$ depends on the critical exponent $\alpha$, giving \cite{TaS2}   
\begin{equation}
n_\mathrm{C}  =  1 + q - \alpha =  q - 1 +3\nu 
\label{nC}
\end{equation}
 where the second equality uses the scaling relation $\alpha = 2 - 3 \nu$.
From the reported value of $n_\mathrm{C}$, the DOS parameter $q$ = -0.50(2) would be obtained from (\ref{nC}) under the assumption that the mobile spinons with $\nu$ = 0.673 are also the main contributor to the specific heat.
Negative $q$ indicates a large concentration of the DOS at low energies, falling away rapidly with increasing energy. 
The obtained $q$ value matches the value -0.5 associated with a quartic dispersion.
The $q$ value differs substantially from $q$~=~1 for a linear Dirac spinon dispersion or $q$~=~0 for a quadratic spinon dispersion or a large area FS state (Fig.\ref{QSLstates}). 
However a quartic spinon state can be excluded, since it would have a large negative $n_\mathrm{\chi}$, which is not consistent with experiment.

\subsubsection{Comparison with models}

The properties of various QSL excitation models are compared with the measured exponents in Table \ref{table1}.
As we have noted, spinon FS and quartic spinon models give a poor match with experiment.
A quadratic spinon dispersion can also be excluded, since both $n_\mathrm{C}$ and $n_\mathrm{\chi}$ are inconsistent with experiment.
In contrast, linear spinon dispersion is much more promising, since both $n_\mathrm{D}$ and $n_\mathrm{\chi}$ are matching with experiment.
The predicted $n_\mathrm{C}$ in this case is close to 2, which is significantly larger than the experimental value.
However, we note that in practice such a large non-phonon exponent would be masked by the phonon contribution with a similarly large exponent \cite{Misumi20}.
It would then be consistent with the experimental data to assign the observed $n_\mathrm{D}$ and $n_\mathrm{\chi}$ to spinons with linear dispersion, 
provided the observed low value of $n_\mathrm{C}$ can be assigned to another type of excitation that dominates the low temperature specific heat.
A natural assignment for this would be the singlet 
to singlet
excitations of a QSL, that do not contribute to the spin transport or magnetic susceptibility.   
If these singlets have the same $\nu$ value as the spinons, then $q$ must be negative, as discussed earlier.
Alternatively a simple quadratic dispersion model for the singlets can be assumed, 
leading to a $\nu$ value close to the mean field value of 0.5 (Table \ref{table1}, line 6), which would seem to be a more reasonable assignment. 
One possible interpretation for these singlets is that they are the vison vortices that would be associated with a Z$_2$ QSL \cite{Wen91,Qi09,Zhou17}.
Following these considerations, our best assignments of QSL excitations against experiment are shown in bold in Table \ref{table1}. 

\subsection{Dimensional reduction}

At this point we return to discuss the orgin of the strong dimensional reduction observed in this system at low temperature.
The dimensional reduction bears some similarity to the reduction to two dimensions reported for the phase boundary near the quantum critical point in the three-dimensional spin-dimer system BaCuSi$_2$O$_6$ \cite{Sebastian06}.
In our case there is no inherent geometrical frustration of the interlayer interactions, but the QSL state can be viewed as an inherently quantum critical phase and the origin of the reduction in this picture is the quantum entanglement.
A simple picture of the origin of the reduced interlayer interaction in the low temperature quantum state can then be derived from consideration of the reduced probability of having an unentangled unpaired spin that can interact with another unpaired spin in the adjacent layer. 
The reduction factor of order 100 found experimentally on entering deep into the QSL region would then correspond to having one in ten such spins in each layer, which is consistent with the entanglement length estimate that we derived in section \ref{Entanglement}. 

\subsection{Conclusions}

In summary, from these results it can be seen that MOF-based kagome systems can have some clear advantages over purely inorganic-based systems such as herbertsmithite in allowing us to measure the intrisic properties of the frustrated $S$=1/2 spins on a highly ideal two-dimensional kagome lattice.
The main advantages here are that the interlayer magnetic interactions are very weak and there are no interlayer sites for the transition metal ions that could act as localized defect spins and obscure the intrinsic properties.

Another beneficial factor here is that the moderate $J$ value provides a convenient energy scale that allows experiments to easily span between the quantum and classical regimes. 
This has allowed the intrinsic magnetic properties of the ideal kagome QSL and its entanglement properties to be elucidated more clearly than was possible in previously studied inorganic kagome systems having larger $J$. 

The present system is effectively gapless, with the upper limit to the gap being of order 0.02~$J$. 
Different QSL models were compared against the experimental data and it was found that a large area spinon FS model is not compatible with experiment.
QSL models with quadratic or quartic dispersion were also found to be at variance with experiment, 
whereas a Z$_2$-linear Dirac QSL with both magnetic and non-magnetic excitations was shown to be consistent with $\mu$SR, magnetic susceptibility and specific heat. 

\vspace{3mm}
The muon raw data files are available from the ISIS facility \cite{data}.  Other parts of the data are available from the corresponding authors upon reasonable request.

\acknowledgements{
The authors acknowledge the financial support from the European Union (ERC AdG Mol-2D 788222, ERC StG 2D-SMARTiES 101042680, FET OPEN SINFONIA 964396 and PATHFINDER 4D-NMR), the Spanish MCIN (2D-HETEROS PID2020-117152RB-100, co-financed by FEDER, and Excellence Unit “María de Maeztu” CEX2019-000919-M) and the Generalitat Valenciana (PROMETEO Program, PO FEDER Program IDIFEDER/2018/061 and grant CIDEXG/2023/1). This study forms part of the Advanced Materials program and was supported by MCIN with funding from European Union NextGenerationEU (PRTR-C17.I1) and by Generalitat Valenciana. Part of this work was carried out at the ISIS Neutron and Muon Source, STFC Rutherford Appleton Laboratory, U.K. The computations were performed on the Tirant III cluster of the Servei d’Informàtica of the University of Valencia and on the SCARF Compute Cluster at STFC.

\vspace{3mm}
EC conceived the project, VGL and MCL prepared and characterised the sample, FLP took the muon data, analysed the results and prepared the first draft of the paper, DLA carried out the DFT calculations supervised by JJB and reported the results. All authors reviewed and refined the manuscript.
}

\section{Appendix}
Density of states plots are shown here for the periodic AB slipped-parallel packing model of section \ref{slipping-section}.

\begin{figure}[htb]
\includegraphics[width=1\columnwidth]{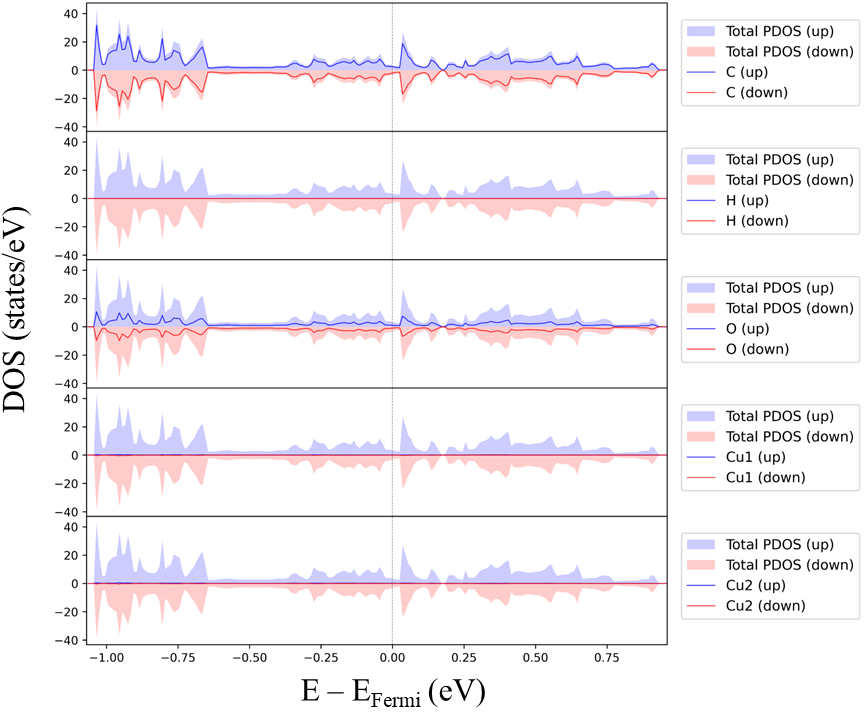}

\vspace{5mm}
\includegraphics[width=1\columnwidth]{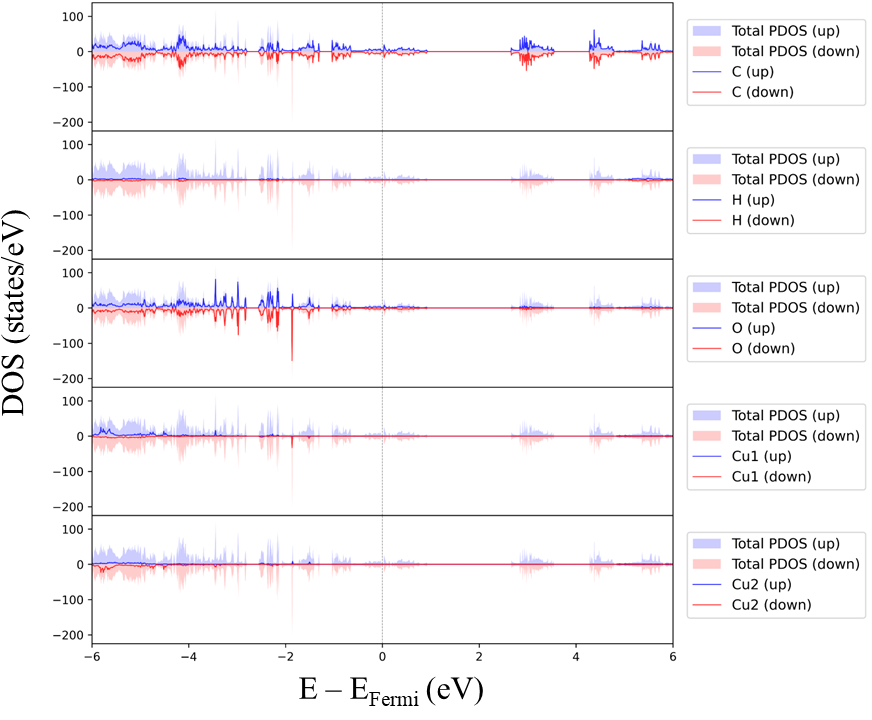}
\caption{
Atom contributions to the density of states for the more stable slipped configuration. Close to the Fermi level (top) and over a wider energy range (bottom).
}
\label{fig-DOS1}
\end{figure}

\begin{figure*}[htb]
\includegraphics[width=1\columnwidth]{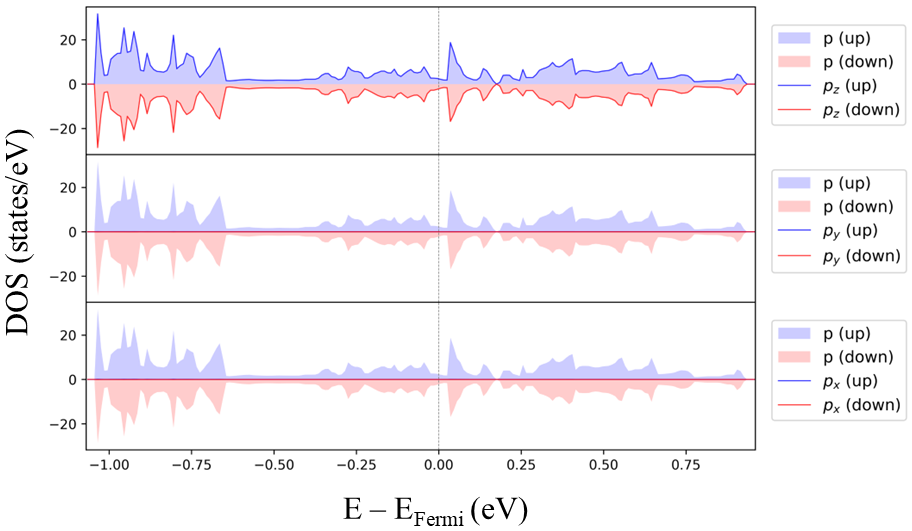} 
\includegraphics[width=1\columnwidth]{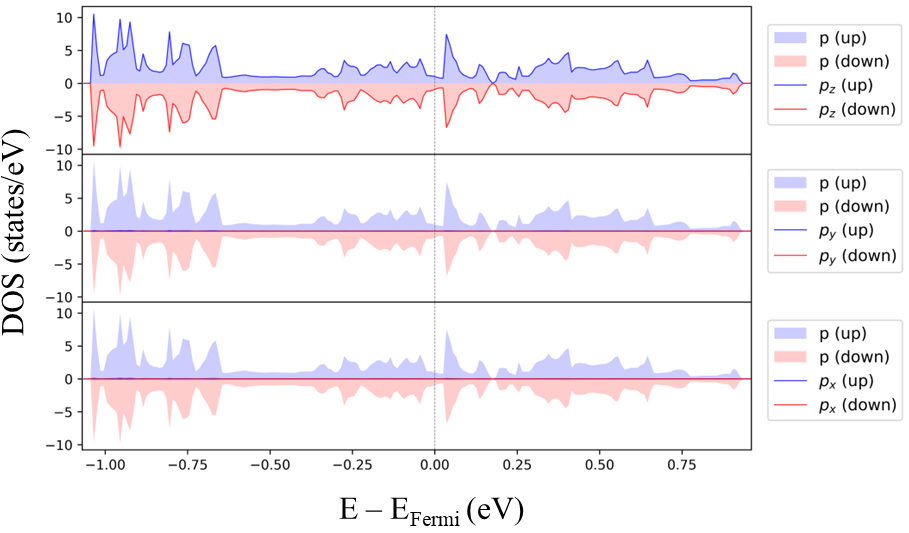}
\caption{
Orbital contributions within the PDOS of Fig. \ref{fig-DOS1} for C atoms (left) and O atoms (right).}
\label{fig-DOS2}
\end{figure*}


\begin{thebibliography}{10}
\bibitem{naming}
The same material is called Cu$_3$(HHTP)$_2$ in some of the literature. We however prefer to use Cu$_3$(HOTP)$_2$, which reflects the deprotonated molecular structure of the HOTP molecule in the compound, in contrast to the original fully protonated HHTP molecule.
\bibitem{data}
The muon data files can be found at DOI:10.5286/ISIS.E.RB2010775.
\end{thebibliography}
\end{document}